\newcounter{myctr}
\def\myitem{\refstepcounter{myctr}\bibfont\noindent\ifnum\themyctr>9\else\phantom{0}\fi\hangindent17pt\themyctr.\enskip}

\documentclass{ws-ijqi_}

\begin{document}

\markboth{A.~Poppe, M.~Peev and O.~Maurhart}{Outline of the SECOQC
Quantum-Key-Distribution Network in Vienna}

\catchline{}{}{}{}{}

\title{OUTLINE OF THE SECOQC\\
QUANTUM-KEY-DISTRIBUTION NETWORK IN VIENNA}

\author{A.~POPPE, M.~PEEV and O.~MAURHART}

\address{on behalf of the Integrated European Project SECOQC -- Development of a\\Global Network for
Secure Communication Based on Quantum Cryptography\footnote{For
affiliations of the SECOQC partners see
\texttt{http://www.secoqc.net/html/partners.html}}.
\\
andreas.poppe@arcs.ac.at}

\maketitle

\begin{abstract}
A Quantum Key Distribution (QKD) network is currently implemented in
Vienna by integrating seven QKD-Link devices that connect five
subsidiaries of SIEMENS Austria. We give an architectural overview
of the network and present the enabling QKD-technologies, as well as
the novel QKD network protocols.
\end{abstract}



\section{Motivation}

The rapid progress in theory\cite{GLLP,cryptodusek,RMP} and
experiment\cite{cryptogisin} of QKD techniques has been reflected by
a number of successful demonstrations in the last few years. Many
groups all over the world have put forward QKD setups operating in
the standard point-to-point modus, thus realizing what we denote as
\textit{QKD-Links} [Fig.~1(a)]. Initial attempts to assemble QKD
based networks have also been launched.\cite{BBN}

A number of research spin-offs (IdQuantique, MagiQ and SmartQuantum)
offer QKD technology but also some internationally operating
companies (e.g. Toshiba, Thales, HP and NEC) develop QKD setups
aiming at a market realization in near future. The secure
communication solutions pursued are typically based on  dedicated
high-end symmetric-encryption devices with frequent key change,
whereby fresh key is constantly generated by QKD devices. The
resulting QKD-Link-Encryptors (with encryption rates potentially up
to 10 Gb per second) offer comparable functionality to existing
products but feature higher level of security. Broad proliferation
of QKD technology is hindered however by a number of road blocks
revolving typically around: the point-to-point paradigm and
correspondingly the quadratic scaling of the initial secrets with
the number of users, the question of integrability in existing
networks, the high price of QKD devices, but also around issues like
missing standards. As a result, QKD appears to be forced into a
narrow niche market.

The EU funded FP6 project SECOQC -- Development of a Global
Network for Secure Communication based on Quantum
Cryptography\footnote{See \texttt{http://www.secoqc.net}. The work
presented here originates from the combined efforts of 41 project
partners working together in various subgroups.} aims to work
against these limitations. A standardization initiative and a
dedicated effort on optimization of single-photon-detection (a
major technological challenge for QKD resulting in a heavy cost
factor) are part of this project. The main goal, however, is
taking in operation and testing an integrated secrets'
distribution network, designed as a Quantum Back Bone network
(QBB) deployed for test purposes in the city of Vienna\footnote{To
be demonstrated in September 2008.}. Thereby heterogenic
QKD-systems by different SECOQC partners are connected with node
modules, which overtake the network functionalities. The QBB,
deployed on a typical telecommunication metropolitan area network
(MAN) is an important step towards demonstrating the feasibility
of fully networked multi-point to multi-point quantum key
distribution. A target market of networked secure communication
solutions for organizations with distributed facilities such as
governmental institutions, companies and banks is addressed.

\begin{figure}[h]
\begin{center}
\includegraphics[height=5cm]{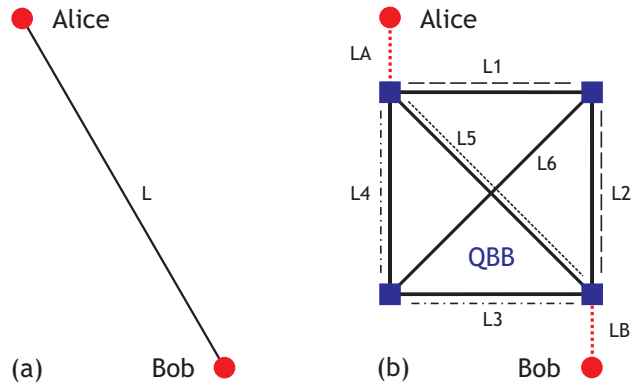}
\caption{(Color online). Alice and Bob want to share a secret key.
(a) Point-to point QKD-Link L: A typical solution nowadays. (b)
Quantum network: The QKD-Links LA and LB are used to access the QBB,
the nodes of the QBB being represented by squares (in blue) and the
corresponding QKD-Links being denoted by L1 through L6. All
QKD-Links provide unconditionally secure channels. An end-to-end
secure channel from Alice to Bob along the paths LA-L5-LB,
LA-L1-L2-LB and LA-L3-L4-LB allows secure transmission of a secret
key.}
\end{center}
\end{figure}
\vspace*{-10pt}

\section{Advantages of Using QKD-Networks}
The underlying idea is to build up a network for distributing
secrets out of single point-to-point QKD-Links. The corresponding
QKD-Link end points (i.e. the QKD devices) are situated in network
nodes. These nodes are secure sites, in which one or more QKD
devices are located together with a central node module, dedicated
to processing, storage and communication. These node modules are the
network agents taking over (in the SECOQC approach) full control of
classical communication channels, the management of the generated
secret keys, their information-theoretical secure transmission from
node to node on a hop-by-hop manner,\cite{all} and issues like
finding paths to distant nodes and ensuring the synchrony of secret
key provision to key-consuming applications across the network.

A QKD-network, following this design, is then an infrastructure for
distributing the secret keys between nodes on a many-to-many basis
over potentially unlimited distances. Under the assumption that the
nodes can be trusted, it allows to utilize the information-theoretic
security of quantum key distribution  and achieve unconditional key
distribution across the network. It should be underlined that while
other approaches to designing quantum key distribution have been put
forward (see e.g.\ Ref.~\refcite{all} for a discussion) the  trusted
repeater architecture, adopted by SECOQC,\cite{peev} is the only
currently feasible one that allows \textit{both} going beyond the
point-to-point and the limited distance constraints of standard QKD.

Figure~1(b) shows an universal building block: a rectangle with
diagonals that is formed by four QBB-nodes (squares) connected
together by six links. Additionally, each QBB node may act as a
possible entry for end-users to be connected to the QBB by a single
QKD-Link. This quantum access network (QAN) is shown in Fig.~1(b)
for an example of two users, Alice and Bob (circles).

In what follows, we list the main advantages of QKD-networks over
single QKD-Links:

\begin{itemlist}
\item To include an additional end-user into the network, only a
single QKD-Link is needed for connecting him to the closest node
of the QBB. Then all other users already attached at the QBB are
potential partners and can share a key with him.

\item For a high number of end-users, the scaling of a QKD-network
comprising a QBB and QANs is strongly favorable. For single
point-to-point connections the number of needed QKD-Links
increases as $\frac{N(N-1)}{2}$ for $N$ users. The presented
network grows only linearly with $N$, as this is the number of the
needed QAN-Links.

\item In a metropolitan area, setting a direct fiber link between
Alice and Bob would typically be far longer than the corresponding
distances connecting both end-users to the QBB. Due to this lower
loss of the access links and the potentially high supply of secure
key through the QBB, the overall key rate will be higher using a
QKD-network.

\item Within the QBB, links can be cascaded in series to connect
users at longer distances and in parallel to obtain higher values
of secure key rate.  Fig.~1.b shows the case of a key request of
user Alice and Bob connected with corresponding links LA and LB to
the QBB. Here link L5 is connected in parallel to a link formed by
L1 and L2 and another one by L3 and L4. Assuming a fixed cost per
QKD device, one can look for an optimum in overall cost as
function of the single-link distance between two nodes. For a
number of advanced QKD-systems, the corresponding optimal distance
turns out to be around 25 km for a one-dimensional chain of links
as well as for a two-dimensional network.\cite{private} This is
also the average
distance in the SECOQC Demonstrator.

\item In the case of an intruder that causes one link to fail due
to any attack (e.g.\ L5 in the upper example), the requested key
could
still be delivered by other remaining routes through the network.

\item Furthermore, with the same argument, if one link is
corrupted by the powerful denial-of-service attack, the pre-shared
initial secrets of the link under attack will run out. In a
single-link scenario, human interaction is needed to restore the
pre-shared secret. In contrast, in a quantum network, this key can
be restored over the remaining quantum links.

\end{itemlist}

\section{Network Demonstration in Vienna 2008}

To introduce and demonstrate new network functionalities and to show
the various advantages over single links, we implement a universal
rectangular building block (Fig.~2) out of four stations in Vienna
(SIE, ERD, GUD, BREIT) and extend it by one node at a nearby city
St. P\"olten. All QBB-Links are implemented including the two
diagonals and two short QKD-connections towards the end-users.

\begin{figure}[h]
\begin{center}
\includegraphics[height=4cm]{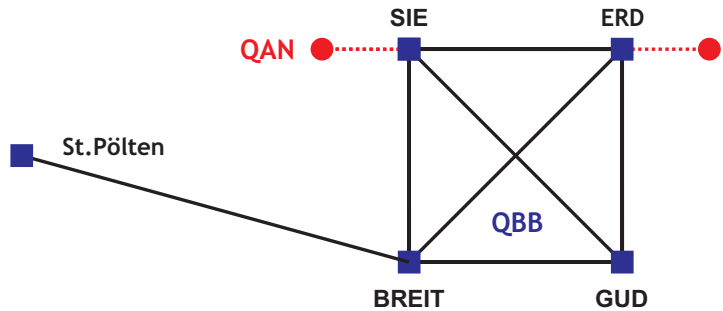}
\vspace*{8pt} \caption{(Color online). Concept of the deployed
universal building block for quantum key distribution networking.
The names refer to the stations of the ring network.}
\end{center}
\end{figure}

In contrast to the architecture of the DARPA quantum
network,\cite{BBN} in which the quantum channel was actively
controlled by a fiber-optical switch depending on the
communication partners, in the presented SECOQC quantum network
each pair of QKD-devices forming a QKD-Link remain together. The
task of each QBB-Link is to grow as much key as possible and to
hand it over to the network node, no matter which end-user will
request for it
afterwards.

All of the used QKD devices have reached high maturity before
leaving the various labs and are being prepared for deployment and
operation at field conditions. The feasibility of the discussed
network architecture will then be tested in real-life conditions.

\subsection{Telecom fiber ring network from Siemens}

All the links will be operated on a typical ring-shaped network to
connect office-buildings of SIEMENS Austria. Fig.~3 shows the
fiber ring with a circumference of approximately 63 km. We use
four node-stations distributed well over the whole city of Vienna
and one in St. Poelten, another city connected by an 85~km long
fiber. The fibers for both cross-connections are also physically
deployed in the ring (connecting the two pairs of non-neighboring
nodes respectively).

\begin{figure}[h]
\begin{center}
\includegraphics[height=7cm]{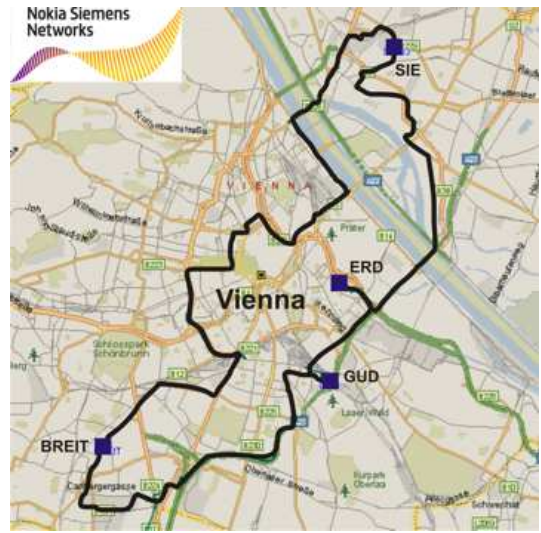}
\vspace*{4pt} \caption{(Color online). City map of Vienna including
the fiber ring network and the stations SIE, ERD, GUD and BREIT.
Another station, St. P\"olten,  is located far outside of Vienna.}
\end{center}
\end{figure}
\vspace*{-10pt}

Each QKD-device pair is connected by their own dark fiber serving
as an exclusive quantum channel. The classical communication is
managed by dedicated node modules using one additional fiber-loop
in the ring, which is realized by a set of communication lasers
multiplexed using CWDM spacing.

\subsection{QKD-link-devices for the Quantum Back Bone QBB}

For a convincing demonstration of the network features, SECOQC has
decided to develop high-rate stable QKD-Link devices. Only devices
with a secure key rate of more than 1 kbit/s after authentication
over 25 km of fibers are accepted for the stage of deployment.
Clearly the link devices must fulfill tight security
standards.\cite{RMP} Moreover, the devices must be autonomously
delivering the key for more than 6 months without human interaction.
The latency time for a new start (for example, after change of the
fiber connection) is limited to 1 minute.

For the stage of deployment the following systems have met these
criteria and form the QBB-network:

\begin{itemlist}

\item IdQuantique (Switzerland): The Swiss company delivers three
devices-pairs usually referred to as ``Plug \& Play''
system.\cite{idq} By attenuating a strong pulse to single-photon
level within one round-trip, no additional compensation of the
drifts from the interferometers and fiber are
needed.

\item GAP Geneva/IdQuantique/ARC (Switzerland/Austria): The
university group headed by Prof. Gisin develops one QKD system in
which the position of non-empty time slots of the attenuated light
pulses carries the transmitted bit value.\cite{COW} Coherence
between the photons reveals any intruder who would try
extracting information out of the system.

\item Toshiba (UK): The group of Andrew Shield develops one QKD
system with two interferometers, which are stabilized by classical
pulses sent after the quantum signals (weak coherent pulses
approximating single photons). BB84 in the phase modulation
version is implemented.\cite{decoy} To beat the photon number
splitting attack for weak coherent pulses, i.e increase the secure
key rate at 25
km, decoy states are implemented.

\item University of Vienna/ARC/KTH (Austria/Sweden):
Entangled photons have been studied by Prof. Zeilinger for a long
time and are now being also transferred to telecommunication
wavelength. Implemented this fundamental quantum-mechanical property
by the BBM92 protocol, the correlations of detections of entangled
photon-pairs at different QKD-devices are used to generate the
key.\cite{entangled}

\item CNRS/Thales/ULB (France/Belgium): The consortium lead by
Prof. Grangier develops the only continuous variable QKD system in
the prototype. Stronger pulses with tens of photons in each pulse
are used instead of weak coherent pulses. Homodyne detection
replaces in this case the usage of single photon
detectors.\cite{CV}

\end{itemlist}

\subsection{QKD-link-devices for the Quantum Access Network (QAN)}

In order to connect an end-user device to the QBB, a single link
towards one QBB-node must be established, whereby the shortest
connection is aspired. For this quantum access network any kind of
QKD devices can be employed in principle. The secure key rate to be
generated for each user is expected to be far below the typical
connection load at the QBB. Therefore QAN-Links call for simple
devices with lower rates, but also of lower cost. As a special case,
to extend QKD-networks to areas with lower penetration of deployed
fiber infrastructure, free space QKD is an appropriate alternative.
For this reason, two free-space QKD systems have been included:

\begin{itemlist}
\item Ludwig-Maximilian-University Munich (Germany): The
free-space system developed by the group Prof. Weinfurter has a
typical working distance of 500 m.\cite{freespace} To reach typical
MAN-like distances the possible distance can be extended to 3 km. Up
to now, only night-time operation is possible, but the used
single-photon detectors based on Silicon permit higher key
generation rates.

\item University of Bristol (UK):
The team under Prof. Rarity developed a portable
QKD-device.\cite{ATM} The operation distance is only a fraction of a
meter, but the aim of this unique development is to enhance the
security of existing applications like the interaction with ATMs by
creating hand-held consumer devices for PIN protection and
authentication.

\end{itemlist}

\subsection{Network nodes}
All end-devices of the QBB-Links and at least one device of each
QAN-Link are situated within a network node (Alice 1, Alice 2 and
Alice 3 in Fig.~4) communicate with the node module. As mentioned
before, each pair of QKD-devices form a link and will remain
together in this presented network architecture. We do not implement
switching on the level of QKD-Link devices, because one goal was to
allow connectivity to the network for a broad range of heterogeneous
QKD-devices only by adopting common interfaces and protocols to the
network node-module.

\begin{figure}[h]
\begin{center}
\includegraphics[height=6.5cm]{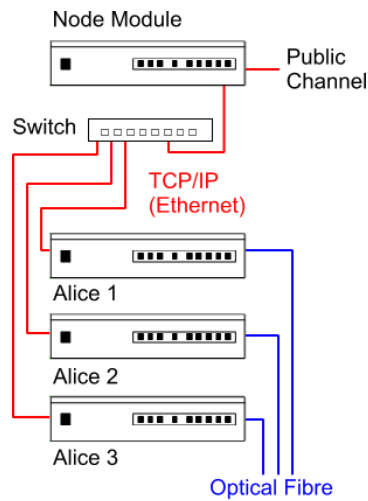}
\caption{(Color online). The node module manages the secret keys and
the classical communication generated by all QKD devices. Device
independent operation is ensured by a common interface. The needed
authentication of messages is also provided by the node module.}
\end{center}
\end{figure}

The task of each QBB-Link is to continuously grow as much key as
possible and to hand it over to the network nodes. Thereby a node
module overtakes the important task of storage and handling of the
generated keys. On one hand, the node module oversees the separate
link-devices in the node, and even more importantly in the presented
architecture, it provides them with classical communication
connectivity to the partner devices of the respective QKD-Link. The
indispensable message authentication of the classical part of the
QKD-Link communication needed to overcome potential
man-in-the-middle attacks is also provided by the node module, the
QKD-protocol stack of each link being responsible to decide which
communication messages are to be authenticated.

On the other hand, the node module is further responsible for the
overall network functionality by executing the QKD higher-level
network protocol stack (discussed in more detail below), which in
charge of routing and transporting the generated secrets on the
network level. The node-module functionality implies that all
node-modules are connected over public channels with each other
(Fig.~5).

\begin{figure}[h]
\begin{center}
\includegraphics[height=12cm]{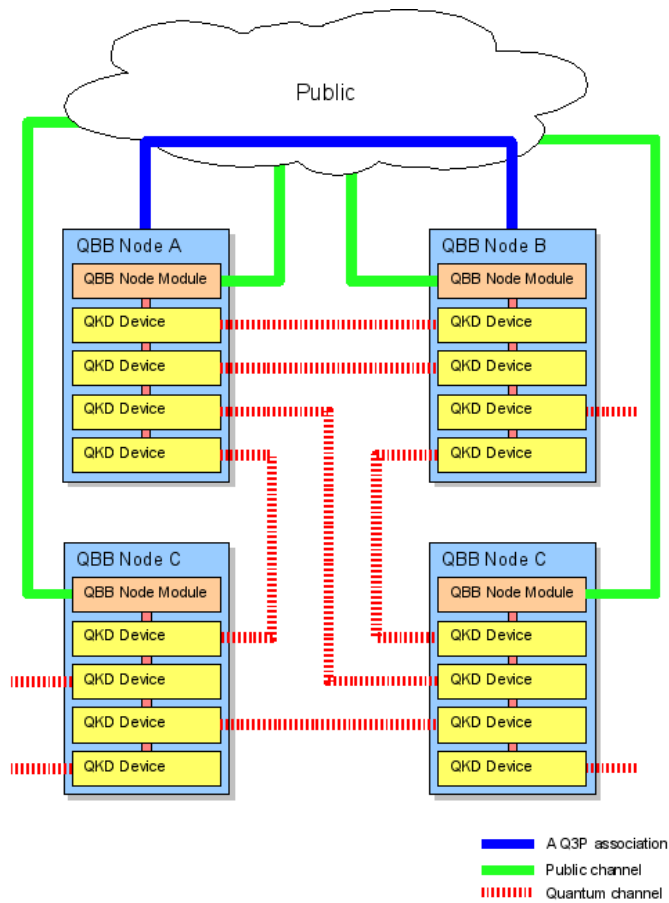}
\caption{(Color online). Four nodes of the QBB are connected with
quantum channels and communicate over the public channels. Between
nodes A and B, one specific connection is highlighted to emphasize
the communication of the Q3P network protocol.}
\end{center}
\end{figure}
\vspace*{-10pt}

\section{Network Protocols}

As  already mentioned, several different devices will compose the
SECOQC QBB network. As each of them has different characteristics
and interfaces, a common protocol has been designed to access
services provided by the devices in an uniform manner: the
Quantum-Point-to-Point-Protocol or Q3P.\cite{maur}

This protocol serves as Point-to-Point Protocol between a pair of
QBB nodes (see Fig.~5) which enables the devices underneath to
carry out the classical communication protocol for key
distillation over an authenticated channel, while simultaneously
providing authentication and encryption for upper layers in the
network. Both authentication and encryption are
information-theoretically secure.

Using Q3P as an uniform building block interconnecting a pair of
QKD devices one can now employ traditional network protocols on
top of it. However most widespread protocols like TCP/IP are not
directly compatible with the specific requirements of quantum key
distribution: network reliability in terms of TCP-like re-sending
packets has to be considered carefully since the content of the
packets in a QKD case is highly sensitive with respect to
confidentiality and/or authenticity.

Therefore well known protocols have been adapted to form a new set
of end-to-end networking mechanisms.\cite{dianat} The QKD Routing
Layer (QKD-RL) Protocol addresses the routing information within
the QBB nodes. This protocol follows the pattern of OSPF but
includes essential modifications to address the specific
requirements arising from the sensitivity and scarceness of key
material.

Finally, the QKD Transport Layer (QKD-TL) Protocol adopts TCP/IP
and introduces new approaches to dealing with highly congested
networks based on quantum key resources. This protocol finally
lets users exchange information across the network with
perfect confidentiality and authenticity on an end-to-end basis.

All three protocols sketched above are independently designed and
each presents a standard-like interface, which can easily be
introduced in current telecom network infrastructures. No
applications running on upper layers need to be modified in order to
use the unconditionally secure key material.

\section{Conclusions and Outlook}

A quantum key distribution network covering a full inner-city
metropolitan area network using seven fiber-based QKD devices,
realized by five different working principles and two free-space
QKD setups will be demonstrated in September 2008 in Vienna. This
implementation in the framework of the EU-funded FP6-project
SECOQC clearly shows the feasibility of constructing highly
integrated QKD-networks. Heterogeneous modern QKD devices are
combined through common interfaces and universal node modules into
unified secret distribution infrastructure. This important step
towards practical deployment offers end-users and network
providers a vision of future secure communication facilities.

\section*{Acknowledgments}

This work was supported by the European Commission through the
integrated project SECOQC (Contract No. IST-2003-506813).


\end{document}